\documentclass[%
aps,
pre,
superscriptaddress,
a4paper,
12pt,
longbibliography,
reprint,
notitlepage,
floatfix
]{revtex4-1}
\usepackage[english]{babel}
\usepackage{amssymb,amsmath,stmaryrd,array}

\usepackage{graphicx}
\usepackage{subfig}

\makeatletter

\newcommand{\sca}[2]{\ensuremath{\bigl({#1}\cdot{#2}\bigr)}}

\newcommand{\ddiv}{\mathop{\rm div}\nolimits}

\newcommand{\mum}{$\mu$m}
 \newcommand{\bs}[1]{\boldsymbol{#1}}
 \newcommand{\vc}[1]{\mathbf{#1}}
 
 \newcommand{\uvc}[1]{\hat{\mathbf{#1}}}
 
 \newcommand{\ind}[1]{\mathrm{#1}}

\newcommand{\dd}{\mathrm{d}}

\makeatother

\begin{document}
\title{Multiple minimum energy paths and scenarios of unwinding
  transitions in chiral nematic liquid crystals} 
%\date{\today}

\author{Semen~S.~Tenishchev}
\email{tenischev.semen@gmail.com}
\affiliation{ITMO University, Kronverkskiy, 49, 197101 St. Petersburg, Russia}
\affiliation{St.\,Petersburg State University, 199034 St. Petersburg, Russia}

\author{Alexei~D.~Kiselev}
\email[Email address: ]{alexei.d.kiselev@gmail.com}
\affiliation{ITMO University, Kronverkskiy, 49, 197101 St. Petersburg, Russia}
\affiliation{St.\,Petersburg State University, 199034 St. Petersburg, Russia}

\author{Aleksei~V.~Ivanov}
\email[Email address: ]{alxvov@gmail.com}

\affiliation{St. Petersburg State University, 199034 St. Petersburg, Russia}
\affiliation{Science Institute and Faculty of Physical Sciences, University of Iceland VR-III,107 Reykjav\'{\i}k, Iceland}

\author{Valery~M.~Uzdin}
\email[Email address: ]{v\_uzdin@mail.ru}
\affiliation{St.\,Petersburg State University, 199034 St. Petersburg, Russia}
\affiliation{ITMO University, Kronverkskiy, 49, 197101 St. Petersburg, Russia}

\begin{abstract}
We apply the minimum energy paths (MEPs) approach to study the helix
unwinding transition in chiral nematic liquid crystals. 
A mechanism of the transition is determined by a MEP passing through a
first order saddle point on the free energy surface. The energy difference between
the saddle point and the initial state gives the energy barrier of the
transition.
Two starting approximations for the paths are used to find the MEPs
representing different transition scenarios: (a)~the director slippage
approximation with in-plane helical structures; and (b)~the anchoring
breaking approximation that involves the structures with profound
out-of-plane director deviations.
 It is shown that, at sufficiently low voltages, the unwinding
 transition is solely governed by the director slippage mechanism with
 the planar saddle point structures.  
 When the applied voltage exceeds its critical value below the
 threshold of the Fr\'eedericksz transition, the additional scenario
 through the anchoring breaking transitions is found to come into
 play. 
 For these transitions, the saddle point structure is characterized by
 out-of-plane deformations localized near the bounding surface. 
  The energy barriers for different  paths of transitions are computed
  as a function of the voltage and the anchoring energy strengths. 
\end{abstract}

% \pacs{%
% 03.65.Vf, 07.60.La,
% 42.25.Hz, 42.25.Ja,
% 78.20.Jq, 42.70.Df
% }
% \keywords{%
% twisting transitions;
% chiral liquid crystals; anchoring breaking;
% }

\maketitle

%%%%%%%%%%%%%%%%%%%%
\section{Introduction}
\label{sec:intro}
%%%%%%%%%%%%%%%%%%%

Helical superstructures naturally arise 
in certain liquids with long-range orientational order
characterizing
the liquid crystalline (mesomorphic) state
that occurs in a temperature range between the liquid and
the solid (crystalline) phases.
In these orientationally ordered liquids
known as the \textit{liquid crystals} (LCs) 
the molecules tend to align along a preferred direction
typically described in terms of the LC director
which is a unit vector $\uvc{n}(\vc{r})$
representing
the locally averaged direction of the LC molecules at a point $\vc{r}$
in the liquid crystalline material~\cite{Gennes:bk:1993,Oswald:bk:2005}.
It is the presence of the LC orientational order
that leads to an optical and electromagnetic anisotropy
which has been extensively exploited
in the nowadays widespread liquid crystal
technology~\cite{Yang:bk:2006,Obayya:bk:2016}. 

Helical twisting patterns
where the director
rotates in a helical fashion about a uniform twist (helical) axis
spontaneously form in unbounded
chiral liquid crystals
and are caused by
the presence of anisotropic molecules with
no mirror plane~---~the so-called chiral molecules.
These patterns thus represent
self-organized soft helical superstructures.

The supramolecular helical architectures
are at the heart of
a unique combination of photonic properties of
chiral nematic liquid crystals,
otherwise referred to
as the \textit{cholesteric liquid crystals} (CLCs).

LCs are known to be responsive materials
that are highly sensitive to external stimuli such as
electromagnetic fields and boundary (anchoring) conditions.
This responsiveness underpins tunability of the helical structures
underlying most of the fascinating device applications of CLCs
and controllable manipulation of the CLC helical
superstructures
presents a challenging problem which is of vital importance for
both fundamental and technological
reasons~\cite{Bisoyi:advmat:2018,Zola:inbk:2018,Ryabchun:adom.2018,Balamurugan:rfp:2016}. 

 An ideal CLC helix is specified by
 orientation of the twisting axis, $\uvc{h}$, and
 the \textit{helix pitch}, $P$
that also govern its optical properties.
In planar confining geometry of typical CLC cells, where CLC is sandwiched
between two parallel bounding surfaces (substrates),
the planar Grandjean  structure (texture)
which is 
the uniform standing helix state
with the helical axis $\uvc{h}=\uvc{z}$ normal to
the substrates
\begin{align}
  \label{eq:clc_helix}
  \uvc{n}=\cos\phi\,\uvc{x}+\sin \phi\,\uvc{y},
  \quad
  \phi=q z +\phi_0,
\end{align}
where $q=\pm 2\pi/P$ is the \textit{helix wave number}
which is positive (negative) for the right (left)-handed helix,
exemplifies the special case
of anisotropic one dimensional (1D) photonic crystals.
It is characterized by a chirally sensitive photonic bandgap.
Circularly polarized light with helicity identical to
the handedness of the helix cannot propagate,
and selective reflection takes place.

The well-established continuum theory describing the phenomenology of CLCs
is formulated in terms of
the Frank-Oseen free energy functional $F_{\ind{el}}[\vc{n}]$
and the elastic free energy density $f_{\ind{el}}$~\cite{Frank:1958,Oseen:1933}
\begin{align}
  \label{eq:frank}
  &
    F_{\ind{el}}[\vc{n}]=\int_V f_{\ind{el}}\,\dd v,
    \quad
    f_{\ind{el}}=\frac{1}{2} \Bigl\{ K_1 ({\nabla}\cdot\vc{n})^2
\notag\\
  &
    +K_2
    \left[ \vc{n}\cdot{\nabla\times\vc{n}}+q_0 \right]^2
    +K_3\,[\vc{n}\times(\nabla\times\vc{n})]^2
    \notag\\
  &
    -K_{24} \ddiv\left(
\vc{n}\ddiv\vc{n}+ \vc{n}\times(\nabla\times\vc{n}) \right)
\Bigr\}\,, 
\end{align}
where $K_1$, $K_2$, $K_3$ and $K_{24} $ are the splay, twist, bend
and saddle-splay Frank elastic constants.
As a manifestation of the chirality caused by the broken mirror symmetry
the expression for the bulk free energy~\eqref{eq:frank}
contains a chiral term proportional
to the equilibrium value of the CLC twist wave number, $q_0$.

The parameter $q_0$
known as the free twist wave number or the free twisting number,
gives the pitch $P_0\equiv 2\pi/|q_0|$ of equilibrium helical structures
in unbounded CLCs.
The pitch $P_0$
depends on the molecular chirality of CLC constituent
mesogens and ranges from hundreds of nanometers to
many microns or more, depending on the system.

An efficient method widely used to prepare
CLCs is doping nematic LC mixtures
with chiral additives that induce a helical structure~\cite{Oswald:bk:2005,Balamurugan:rfp:2016}.
For photosensitive chiral dopants (photoswitches),
their helical twisting power and thus
the CLC equilibrium helix pitch $P_0$
can be controlled by light
through photoinduced changes in 
chiral molecular switch conformation that influence
the LC's helical twisting
power~\cite{Vinograd:mclc:1990,Delden:chem:2003,Eelkema:lc:2011,Katsonis:jmatchem:2012,Kiselev:pre:2014,Zheng:nature:2016,Bisoyi:anie:2015,Huang:marc:2017}. 
Phototunability of the helix pitch
leads to a variety of technologically promising effects such as the
phototunable selective reflection,
i.e. a light-induced change in the spectral position of the bandgap~\cite{Kurihara:apl:1998,White:jap:2010,Kosa:nature:2012,Vernon:optexp:2013}.

An important point is that, director configurations in the planar CLC cells
 are strongly affected by the
anchoring conditions at the substrates.
These conditions break the translational symmetry along the twisting
axis and, in general, the helical form of the director field will be distorted.
Nevertheless, when the anchoring conditions are planar and
out-of-plane deviations of the director are suppressed, it might
be expected that the configurations still have the form of the
ideal helical structure~\eqref{eq:clc_helix}.
But, by contrast with  the case of
unbounded CLCs, the helix twist wave number $q$ will now differ
from $q_0$.

A mismatch between 
the twist imposed by the boundary conditions
and the equilibrium pitch $P_0$ may produce two
metastable twist states that are degenerate in energy and can be
switched either way by applying an electric
field~\cite{Berrem:jap:1981}.
More generally metastable twist states in CLC cells appear
as a result of competing influences of the bulk and the surface
contributions to the free energy
leading to frustration~\cite{Kamien:jcmp:2000,Oswald:bk:2005}
and giving rise to multiple local minima
of the energy~\cite{Mottram:cmt:1997}.
Properties of the metastable helical structures are determined by 
the free twisting number $q_0$ and the anchoring energy. 
Variations in $q_0$ will
affect the twisting wave number, $q$, and may
result in sharp transitions~---~the so-called \textit{pitch transitions}~--~between 
different branches of metastable states.

In particular, these transitions 
manifest themselves in
a jump-like temperature dependence of selective
light transmission 
spectra~\cite{Zink:jetp:1997,Gandhi:pre:1998,Zink:1999,Yoon:lc:2006}.
Different mechanisms behind the temperature variations of the pitch in CLC cells and hysteresis phenomena
were discussed in Refs.~\cite{Bel:eng:2000,Bel:eng:2003,Palto:eng:2002}.
A comprehensive stability analysis
of the planar helical structures  in CLC cells with
symmetric and asymmetric boundary conditions
was performed in Ref.~\cite{Kiselev:pre-1:2005}.
The effects of bistable surface anchoring and mechanical strain
on the pitch transitions
have been studied theoretically in the recent papers~\cite{McKay:epje:2012}
and~\cite{Lelidis:pre:2013,Barbero:jmolliq:2018}, respectively.

Another standard and widely exploited technique
to manipulate the helical structures 
uses their sensitivity to
external (magnetic or electric) fields
applied to CLC cells.
An external field will generally
distort
the free energy landscape.
These distortions
lead to a variety of
field-induced orientational effects
such as
the Fr\'eedericksz and unwinding transitions
that have been
attracted considerable attention
in the context of electro-optics of LC display devices.

In the technologically important geometry where
the electric field is applied across the CLC cell,
these effects crucially depend on
a number of factors such as
the cell thickness $L$, the pitch $P_0$,
the applied voltage $U$, the anchoring conditions,
elastic and dielectric properties
of the CLC
material~\cite{Meyer:apl:1968,Berrem:jap:1981,Becker:jap:1985,Schiller:phtrans:1990,Hirning:jap:1991,Smalyukh:pre:2005,Choi:advmat:2009,Valkov:pre:2013}.
In this geometry,
the CLC cell with planar (homogeneous)
anchoring conditions is subjected to
the field applied along the twisting axis
of the planar helical structure~\eqref{eq:clc_helix}.
This is the case of our primary concern.

A systematic theoretical analysis of
electrically induced instabilities
of the CLC helix
underlying the Fr\'eedericksz transition
performed in Ref.~\cite{Valkov:pre:2013}
have shown that the transition
can be either continuous
(second order) or
discontinuous (first order)
depending on the elastic constants,
the pitch, the anchoring energy strengths
and the cell gap.
It is also known that
destabilization of the helical structure
may alternatively occur
through the Helfrich-Hurault mechanism~\cite{Helfrich:apl:1970,Helfrich:jcp:1971,Hurault:jcp:1973,Nemati:jap:2012}
giving rise to undulated patterns
with in-plane electrically dependent periodicity.
These patterns also known as
striped textures or Helfrich distortions
have aroused considerable interest
as switchable CLC diffraction
gratings~\cite{Subacius:apl:1:1997,Subacius:apl:2:1997,Senyuk:ol:2005,Gvozdovskyy:lc:2017,Shtykov:lc:2018,Zola:advmat:2019}.

The purpose of this paper is to explore
global properties of
the free energy landscape
related to transitions
between CLC states
which are local minima (minimizers)
of the free energy.
The free energy pathways
connecting pairs of metastable helical states
appear as basic elements of
a natural mathematical language dealing with
the relevant geometry of
the landscape viewed as a multidimensional
free energy surface~\cite{Kusum:softmat:2015,Bessarab:cpc:2015,Ivanov:pre:2016}.

The key elements associated with the transitions
are the \textit{minimum energy paths} (MEPs) between the initial and
final states on the free energy surface.
Every point on such a pathway
is a free  energy minimum in all but a certain direction in the
configuration space of CLC director structures. 
Maximum along
the MEP determines the \textit{transition state} which is a saddle
point on the free energy surface. 
The MEP itself represents a path with the maximal statistical weight
and defines a scenario of the most probable transition between
the states. 
The energy barrier separating the states can be found as
the difference between the saddle point energy and the energy of the initial state.
When the transition goes through several metastable states,
MEP gives a sequence of the barriers to be passed
in the course of the transition. 

Information about
such energy barriers is required
to asses the effect of thermally activated transitions within
the framework of the rate theory~\cite{Hanggi:rmp:1990,Coffey:advchem:2007}.
Similarly,
in Refs.~\cite{Goldbart:prl:1990,Goldbart:mclc:1991},
the barrier heights and
the Arrhenius formula were
employed
to estimate
the rate of transitions between metastable twist states
and
the effective intrinsic torsional viscosity
of LC cells with strong anchoring conditions.

In this paper we restrict our analysis to the case
where stability of the CLC helix is determined
by the threshold voltages
of the Fr\'eedericksz transition 
and apply
the geodesic nudged elastic band (GNEB) method~\cite{Bessarab:cpc:2015} to
calculate MEPs of
the helix unwinding transition between a metastable CLC twist state
and the untwisted ground state.
From the computed MEPs we identify
two different scenarios for
unwinding of the CLC helix to occur.
These are:
(a) the transitions dominated by in-plane director slippage;
and (b) the transitions involving localized anchoring breaking.
We shall study
how these transitions and their free energy barriers
depend on
the electric field and the anchoring conditions.

The layout of the paper is as follows.
General relations
that determine the characteristics of
the helical structures in CLC cells are given
in Sec.~\ref{sec:theory}.
Then in Sec.~\ref{sec:results} we outline the
numerical procedure that we employ to
compute MEPs and describe the results
obtained using 
the director slippage and anchoring breaking
approximations
as starting approximations for the MEPs.
Finally, in Sec.~\ref{sec:conclusion}
we discuss our results and
make some concluding remarks.

%%%%%%%%%%%%%%%%
\section{Free energy}
\label{sec:theory}
%%%%%%%%%%%%%%%%

We consider a CLC cell of thickness $L$ sandwiched
between two parallel plates that are normal to
the $z$-axis:  $z=-L/2$ (lower substrate) and
$z=L/2$ (upper substrate).
Anchoring conditions at both substrates are planar
(homogeneous)
with the preferred orientation of CLC molecules at the lower and upper
plates defined by
the two vectors of easy orientation: $\uvc{e}_{-}$
and $\uvc{e}_{+}$, where a hat will indicate unit vectors.
These vectors are given by
\begin{align}
  \label{eq:clc-anch-vec}
  \uvc{e}_{\pm}=\cos\psi_{\pm}\,\uvc{x}+\sin\psi_{\pm}\,\uvc{y},
%  \quad 
\end{align}
Then $\Delta\psi=\psi_{+}-\psi_{-}$, is the twist angle imposed by the boundary conditions.

We shall also 
write the CLC free energy functional as a sum of the bulk and
surface contributions:
\begin{align}
  &
  F[\vc{n},\vc{E}]=F_b[\vc{n},\vc{E}]+F_s[\vc{n}],
    \notag
  \\
  &
  F_s[\vc{n}]=\sum_{\nu=\pm 1}
\int_{z=\nu L/2} W_{\nu}(\vc{n})\,\dd s,
\label{eq:f-gen}
\end{align}
where $\vc{E}$ is the electric field,
and assume that both the polar and the azimuthal
contributions to the anchoring energy $W_{\nu}(\vc{n})$
can be taken in the form of Rapini-Papoular potential~\cite{Rap:1969}:
\begin{align}
&
W_{\nu}(\vc{n})=
\frac{W^{(\nu)}_{\phi}}{2}
\Bigl[1-\sca{\vc{n}}{\uvc{e}_{\nu}}^2\Bigr]_{z=\nu L/2}
\notag
\\
&
+
\frac{W^{(\nu)}_{\theta}-W^{(\nu)}_{\phi}}{2}
\sca{\vc{n}}{\uvc{z}}^2\Bigr|_{z=\nu L/2},
  \label{eq:f_s-gen}
\end{align}
where $W^{(+)}_{\phi}$ ($W^{(-)}_{\phi}$) and $W^{(+)}_{\theta}$ ($W^{(-)}_{\theta}$) are the azimuthal
and the polar anchoring strengths
at the upper (lower) substrate.

Then we express the CLC director $\vc{n}$
in terms of the polar and the azimuthal angles,
$\theta$ and $\phi$, as follows
\begin{align}
  \label{eq:n-angles}
\vc{n} = \sin\theta \cos\phi\,\uvc{x}+ \sin\theta \sin\phi\,\uvc{y} + \cos\theta\,\uvc{z},  
\end{align}
where the angles are functions of $z$, $\theta=\theta(z)$
and $\phi=\phi(z)$,
provided
invariance with respect to in-plane translations is unbroken.
After substituting the director parametrization~\eqref{eq:n-angles}
into Eq.~\eqref{eq:f_s-gen}, we have 
the surface potential in the following form:
\begin{align}
  \label{eq:anchoring-pot}
  &
  F_s[\vc{n}]/A=\sum_{\nu=\pm 1}
  \biggl[
    \frac{W^{(\nu)}_{\phi}}{2}\sin^2\theta_\nu\sin^2(\phi_\nu-\psi_\nu)
    \notag
  \\
  &
    +
  \frac{W^{(\nu)}_{\theta}}{2}\cos^2\theta_\nu
  \biggr],
\end{align}
where
$A$ is the area of the substrates;
$\theta_\nu\equiv\theta(\nu L/2)$
and $\phi_\nu\equiv\phi(\nu L/2)$.

\begin{figure*}[!htb]
   \centering
   \resizebox{120mm}{!}{\includegraphics*{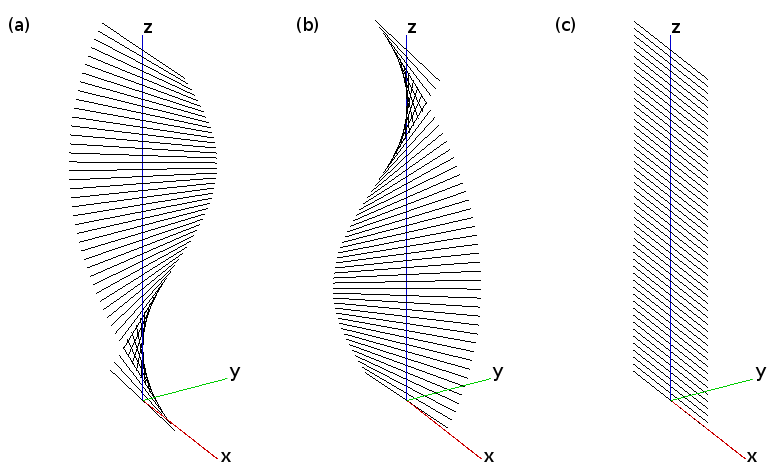}}
   \caption{Field free states: (a)~the metastable left-handed helical (twisted) structure;
     (b)~the metastable right-handed helical (twisted) structure;
     (c)~the stable untwisted (nematic) structure.
     }
\label{fig:stable_states}
\end{figure*}

The bulk part of the free energy functional~\eqref{eq:f-gen}
\begin{align}
  \label{eq:bulk-energy}
  F_b[\vc{n},\vc{E}]=F_{\ind{el}}[\vc{n}]+F_{\ind{E}}[\vc{n},\vc{E}]
\end{align}
is a sum of the Frank-Oseen elastic energy
$F_{\ind{el}}[\vc{n}]$ given by Eq.~\eqref{eq:frank}
and the electrostatic energy of interaction between
the electric field $\vc{E}$ and CLC molecules,
$F_{\ind{E}}[\vc{n},\vc{E}]$.
For the CLC director~\eqref{eq:n-angles},
the elastic energy~\eqref{eq:frank} takes the following form:
\begin{align}
  \label{eq:elastic-energy}
  &
  F_{\ind{el}}[\vc{n}]/A = \frac{1}{2} \int_{-L/2}^{L/2}
  \bigl\{
    K_1(\theta)[\theta']^2 +  K_2(\theta)\sin^2\theta\, [\phi']^2
    \notag
  \\
  &
    - 2C(\theta)\phi'
  \bigr\}\dd z,
  \\
  &
  \label{eq:Ki-eff}
  K_i(\theta) = K_{i}\sin^2\theta+K_{3}\cos^2\theta,
  \: C(\theta) = q_0K_{2}\sin^2\theta,  
\end{align}
where prime stands for derivative with respect to $z$.

In our case, the electric field
is normal to the substrates
$\vc{E}=E_z(z)\uvc{z}$
with $E_z(z)=-V'(z)$,
where $V(z)$ is the electrostatic potential.
It meets the Maxwell equation:
$\ddiv \vc{D}=0$ for the electric displacement $\vc{D}=\bs{\epsilon}\,
\vc{E}$ linearly related to $\vc{E}$ through
the uniaxially anisotropic dielectric tensor $\bs{\epsilon}$
with the components
\begin{align}
  \label{eq:diel-tensor}
  \epsilon_{ij}=\epsilon_{\perp}\delta_{ij}+\epsilon_a n_in_j,\quad
  \epsilon_a=\epsilon_{\parallel}-\epsilon_{\perp},
\end{align}
where
$\delta_{ij}$ is the Kronecker symbol and $i,j\in\{x,y,z\}$;
$\epsilon_{\perp}$ and
$\epsilon_{\parallel}$ are the dielectric constants
giving the principal values of $\bs{\epsilon}$.
So, the normal component of $\vc{D}$,
$D_z=\epsilon_{zz}E_z$ is independent of $z$
and we obtain
the relation linking the voltage applied across the CLC
and $D_z$
\begin{align}
  \label{eq:voltage}
  &
    U=V(-L/2)-V(L/2)=\int_{-L/2}^{L/2} E_z\dd z
    \notag
  \\
  &
    =
  D_z\int_{-L/2}^{L/2} \frac{\dd z}{\epsilon_{zz}(\theta)},
\end{align}
where $\epsilon_{zz}(\theta)=\epsilon_{\perp}+\epsilon_a \cos^2\theta$.
When the applied voltage is fixed,
the electrostatic part of the energy
\begin{align}
  \label{eq:el-en-gen}
  F_{\ind{E}}=-\frac{1}{2}\int_V\sca{\vc{E}}{\vc{D}}\dd v
\end{align}
assumes the form of nonlocal functional:
\begin{align}
  \label{eq:el-en}
  F_{\ind{E}}/A=-\frac{U^2}{2 E[\theta]},
  \quad
  E[\theta]=\int_{-L/2}^{L/2} \frac{\dd z}{\epsilon_{zz}(\theta)}.
\end{align}

In our subsequent calculations,
we shall use the Frank elastic constants
typical for 5CB~\cite{Bogi:lc:2001}: 
$K_{1}=4.5$~pN, $K_{2}=3.0$~pN, $K_{3}=6.0$~pN
and consider the case of the weakly twisted
symmetric CLC cell of the thickness
$L=5$~\mum\  with $q_0L=0.05$,
$W_{\phi}^{(\pm)}\equiv W_{\phi}=0.05$~mJ/m$^2$
and $\uvc{e}_{\pm}=\uvc{x}$.
These parameters are used
to obtain the estimates we briefly discuss below.

In the absence of electric field,
the field free planar CLC helical structures~\eqref{eq:clc_helix}
can be analyzed using the results
of Ref.~\cite{Kiselev:pre-1:2005}.
According to this analysis,
the transitions of our primary concern
involve three (meta)stable structures
shown in Fig.~\ref{fig:stable_states}:
(a)~the left-handed twisted structure with
$qL\approx -3.066$;
and (b)~the right-handed twisted structure with
$qL\approx 3.069$
and (c)~nearly untwisted (nematic) structure with $qL\approx 0.001$;
Since the free twist parameter $q_0L$ is small,
the twisted structures are nearly degenerate in energy.
The nematic structure is the stable state with the lowest
energy (the energy difference per unit area
can be estimated at about $3.0$~$\mu$J/m$^2$).

The threshold voltage of
the Fr\'eedericksz transition $U_{\ind{th}}$
can be estimated
using the analytical relations derived in
Ref.~\cite{Valkov:pre:2013}: $U_{\ind{th}}\approx 0.578$~V.
These relations also predict that, in our case,
the transition leading to instability of the ground state
will be continuous.

\begin{figure*}[!tbh]
   \centering
   \resizebox{80mm}{!}{\includegraphics*{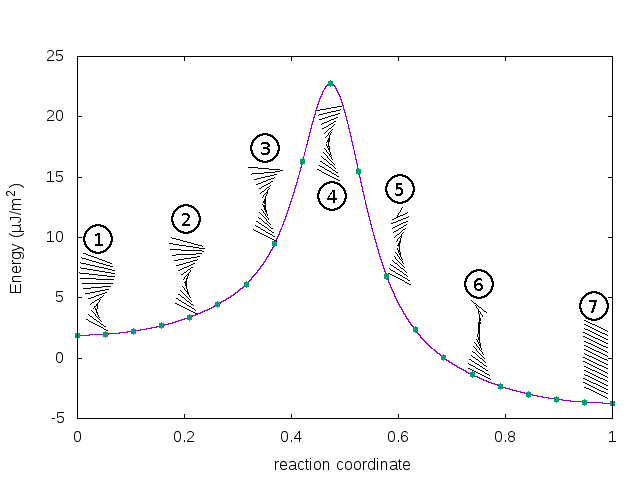}}
   \caption{
     Energy per unit area along the MEP for the director slippage  transition
     computed at $U=0.53$~V and $r_W=1.0$.
     The filled circles correspond to the images of the system
used in the GNEB calculation. The reaction coordinate is defined as
the displacement along the path normalized by its total length.
}
\label{fig:energy_mep_rotate}
\end{figure*}

%%%%%%%%%%%%
\section{Results}
\label{sec:results}
%%%%%%%%%%%%

In this section, we will focus on the unwinding transition
from the metastable left-handed helix state
to the stable nematic (equilibrium) state (see
Fig.~\ref{fig:stable_states}).
Such a situation may occur
in the light-induced pitch
transitions in CLCs doped with photosensitive chiral
dopants~\cite{Kiselev:pre:2014}
when under the action of irradiation
the free twisting number $q_0$ changes
from the initial value close to $-\pi/L$
to the value close to zero.

The states differ in the parity of
half-turns~\cite{Kiselev:pre-1:2005}
and are thus topologically distinct. The latter implies that,
in the strong anchoring limit,
the helical state cannot be smoothly deformed
into the untwisted state without destroying
the local degree of molecular ordering.
By contrast,
in the weak anchoring regime where
the anchoring strength $W_{\phi}$ is not infinitely large,
these states are local minima
of the multidimensional free energy surface
separated by the finite energy barriers.
We shall apply
the minimum energy path approach to study
how these barriers are affected by
the applied electric voltage $U$
and the \textit{anchoring strength ratio}
$r_W=W_{\theta}/W_{\phi}$,
where
$W_{\theta}=W_{\theta}^{\pm}$. 
After brief discussion of our method,
we present the results for two
classes of the pathways representing
the two different scenarios of the unwinding transition:
the \textit{director slippage transitions}
where the out-of-plane deviations of the director
are suppressed ($\theta\approx \pi/2$)
and
the \textit{anchoring breaking transitions}
that involve the states characterized by
profound variations in the
polar angle $\theta$.

\begin{figure*}[!tbh]
   \centering
   \resizebox{80mm}{!}{\includegraphics*{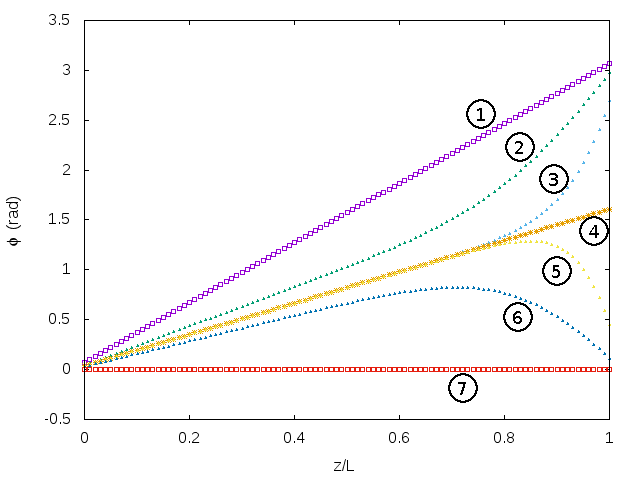}}
\caption{Profiles of the azimuthal angle for each image in the
  MEP for the director slippage transition.
}
\label{fig:rotate_phi}
\end{figure*}

%%%%%%%%%%%%%%%%%
\subsection{Minimum energy paths}
\label{subsec:meps}
%%%%%%%%%%%%%%%%%

In our calculations, the cell is divided into $100$ equidistant layers
and the director orientation is assumed be constant inside each
layer. Since the director in each layer has two degrees of freedom (the
azimuthal and the polar angles), the dimension of the free energy
surface equals twice the number of the layers. This surface can be
regarded as a $200$-dimensional Riemannian manifold defined as a
direct product of $100$ two-dimensional spheres. 

As in Ref.~\cite{Ivanov:pre:2016}, we have used
the
geodesic nudged elastic band
(GNEB) method to find MEPs between local minima on the curved
manifolds~\cite{Bessarab:cpc:2015}. 
This approach involves taking an initial guess of a path
between the two minima and systematically bringing that to
the nearest MEP. A path is represented by a discrete chain
of states, or ``images'', of the system, where the first and the
last image are placed at the local energy minima corresponding
to the initial and final metastable configurations.
In order to distribute the images evenly along the path, springs 
are introduced between adjacent images.
At each image, a local tangent to the path needs to be
estimated, and the force guiding the images towards the nearest
MEP is defined as the sum of the transverse component of the
energy antigradient plus the component of the spring
force along the tangent to the path. The position of intermediate images
is then adjusted so as to zero the GNEB forces.

An important point is that
the MEP connecting the metastable structures
generally depends on the starting approximation for the path.
Variations in the initial approximations may produce
different MEPs. Such paths represent distinct scenarios
of the transition.
Therefore, we have used different initial paths
in order to study the two scenarios of the unwinding transition.
 
To find the director distributions for the initial and final states
we have started from the initial approximation
for the director structure and then
minimized the energy using the velocity projection algorithm~\cite{Jonsson:inbk:1998}.
The position of the maximum (saddle point) along
the MEP  was found using Climbing Image algorithm~\cite{Bessarab:cpc:2015}.

\begin{figure*}[!tbh]
  \centering
    \subfloat[Free energy surface]{
   \resizebox{85mm}{!}{\includegraphics*{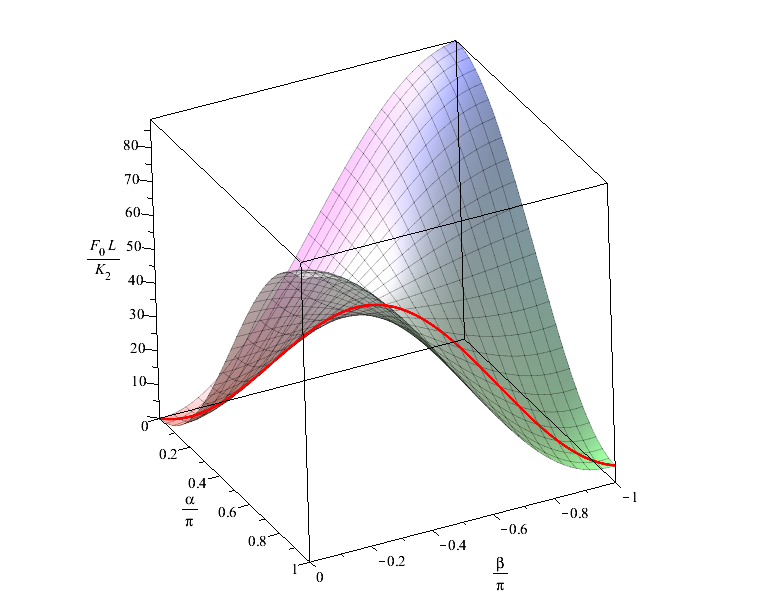}}
   \label{fig:energy_surface_id}
}
   \subfloat[Energy along the saddle-point path]{
   \resizebox{75mm}{!}{\includegraphics*{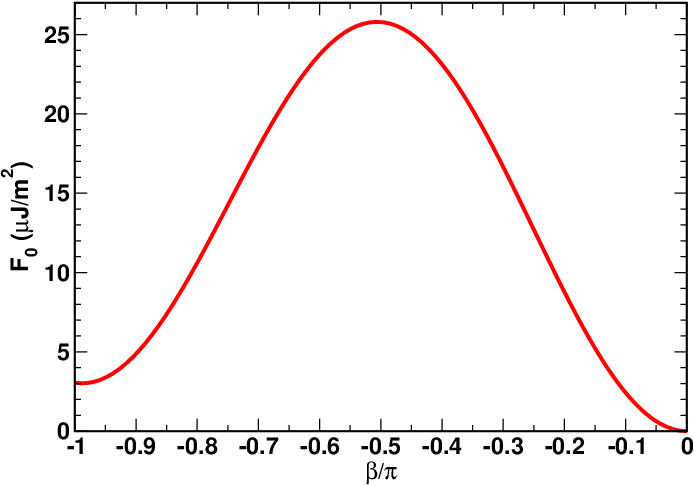}}
   \label{fig:saddle_point_id}
}
  \caption{(a)~Free energy surface for the helical
    structures~\eqref{eq:clc_helix}.
    (b)~Energy along the path passing through the saddle point.
  The energy barrier is $\Delta E\approx 22.6$~$\mu$J/m$^2$.} 
  \label{fig:energy_path_id}
\end{figure*}

%%%%%%%%%%%%%%%%%%%%%%
\subsection{Director slippage transitions}
\label{subsec:direct-slipp-trans}
%%%%%%%%%%%%%%%%%%%%

According to Ref.~\cite{Kiselev:pre-1:2005},
under certain conditions
the pitch transitions being solely governed by
in-plane director fluctuations
do not involve tilted
configurations and
the transition mechanism can be described as director slippage
through the energy barriers formed by the surface potentials.
In what follows, such transitions will be referred to as
the director slippage transitions.

\begin{figure*}[!tbh]
  \centering
  \resizebox{100mm}{!}{\includegraphics*{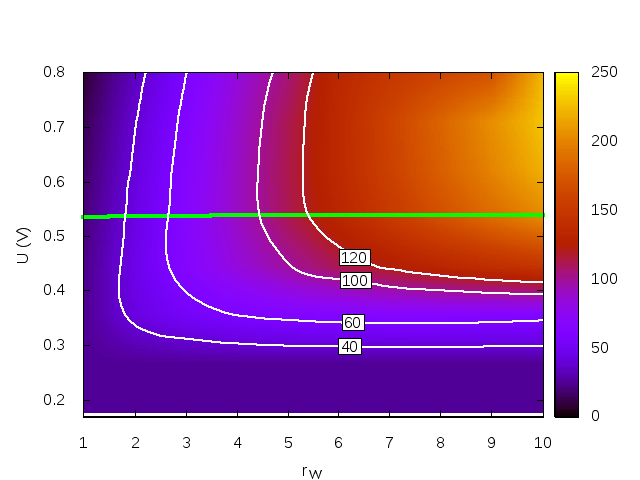}}
  \caption{Energy barrier  map in the $r_W-U$ plane for the director slippage transitions.
    The green line indicates the threshold voltage of
    the Fr\'eedericksz transition $U_{th}$ and the energy barrier scale is
    given in $\mu$J/m$^2$.} 
  \label{fig:barrier_map_rotate}
\end{figure*}

In our calculations,
the left-handed helix
(see Fig.~\ref{fig:stable_states}a)
and the nematic equilibrium structure
(see Fig.~\ref{fig:stable_states}c)
are used as the initial and final states, respectively.
The starting approximation for the path involves
the in-plane helical structures
with $\theta=\pi/2$ assuming
that
the azimuthal angle, $\phi_{+}$, at the top substrate
uniformly varies along the path
approaching the untwisted state value
equal to
the twist angle at the bottom substrate
$\phi_{-}=0$.
In what follows such an initial approximation
of the MEPs will be called the \textit{director slippage
approximation}.

In Fig.~\ref{fig:energy_mep_rotate}, we present
the energies of director configurations along the  MEP computed at
$U=0.53$~V and $r_W\equiv W_{\theta}/W_{\phi}=1.0$.
In addition, figure~\ref{fig:energy_mep_rotate} shows
the director structures for a set of the selected images
(the total number of the computed images is 18)
along the path.
The first image is the metastable
helical structure, whereas the last image is the stable nematic state.
The fourth numbered image is the transition (saddle-point) state
giving the energy that determines the energy barrier
(activation energy) of the transition.
The profiles of the azimuthal angle $\phi$ for CLC structures
shown in Fig.~\ref{fig:energy_mep_rotate}
are plotted in Fig.~\ref{fig:rotate_phi}.
Clearly, similar to the initial and final states,
the profile for the transition state appears to be linear
with the director at the top substrate
oriented along the normal to the easy axis.
By contrast, it turns out that the profiles of other states
along the MEP demonstrate nonlinear
behavior of the azimuthal angle evaluated as a function
of $z$.
Such profiles significantly differ from the Grandjean
texture~\eqref{eq:clc_helix}.

The director slippage scenario implies that,
below the threshold voltage, tilted structures have no effect on
unwinding of the helix and the corresponding MEPs.
For the planar structures with $\theta=\pi/2$, 
the orientation dependent part of
the energy~\eqref{eq:f-gen}
is independent of both the polar anchoring strength
$W_{\theta}$ and the applied voltage $U$.
In particular, 
for the ideal helical structures~\eqref{eq:clc_helix},
this energy is
\begin{align}
  \label{eq:F_0_id}
  &
  F_0(\alpha,\beta)=\frac{K_2}{2 L}(\beta-\beta_0)+\frac{W_{\phi}}{4}
    [2-\cos(\beta+\alpha)
    \notag
  \\
    &
    -\cos(\beta-\alpha)],
\end{align}
where $\beta=q L$ and $\alpha=2\phi_0$,
gives the two-dimensional (2D) free energy surface.
Part of this surface
with the two minima representing
the twisted ($\beta\approx-\pi$) and unwound ($\beta\approx 0$) structures under consideration
is depicted in Fig.~\ref{fig:energy_surface_id}.

The energy along the path connecting the minima and passing through
the saddle point (this path is shown as the red line in Fig.~\ref{fig:energy_surface_id})
is plotted versus the twisting parameter $\beta$ in Fig.~\ref{fig:saddle_point_id}.
It can be readily found that the saddle-point state is
the structure with $\alpha=-\beta=\pi/2$ and the energy barrier
can be estimated at about
$\Delta E=F_0(\pi/2,-\pi/2)-F_0(\pi,-3.07)\approx 22.6$~$\mu$J/m$^2$.
This estimate is close to $W_{\phi}/2$.

When using the minimum energy paths method, 
a MEP evaluated at certain values of
the anchoring ratio $r_W$
and
the applied voltage $U$
gives the energy barrier $\Delta E(r_W,U)$
computed as the difference between
the energies
of the transition (saddle point) state and of the initial twisted
state.
The energy barrier map
in the $r_W-U$ plane
computed for the director slippage transitions
is presented in Fig.~\ref{fig:barrier_map_rotate}.
In agreement with the above discussion,
below the critical voltage $U_{\ind{th}}$,
the director structure of the transition state is uniformly twisted
and the energy barrier is independent of
both the anchoring ratio $r_W$ and the voltage $U$
(for illustrative purposes,
the low voltage part of the map shown in Fig.~\ref{fig:barrier_map_rotate}
is truncated).
In this low-voltage regime,
the above analysis of the ideal helical structures
and the MEP method
give the identical results for the value of the energy barrier.

\begin{figure*}[!tbh]
   \centering
   \resizebox{80mm}{!}{\includegraphics*{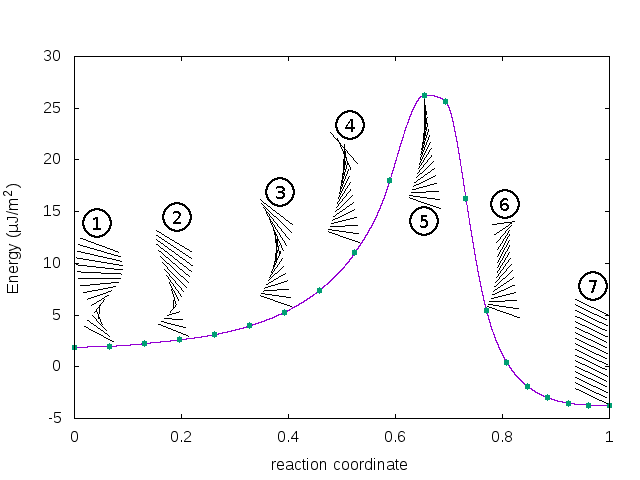}}
   \caption{
  Energy per unit area along the MEP for the anchoring  breaking transition
     computed at $U=0.53$~V and $r_W=1.25$.
   }
\label{fig:energy_mep_up}
\end{figure*}

As is seen from Fig.~\ref{fig:barrier_map_rotate},
above the threshold voltage $U_{\ind{th}}$ of the Fr\'eedericksz transition,
the energy barrier increases with the anchoring ratio
$r_W$ and, in  general, is a nonmonotonic function of
the applied voltage $U$.
In this region,
the planar states are unstable
and all the structures involved in the transitions
are deformed by the applied field.
These are the tilted structures characterized by
profound out-of-plane director deviations.
The MEPs above the critical voltage $U_{\ind{th}}$
will be discussed later on in Sec.~\ref{subsec:crossover-trans}.

\begin{figure*}[!tbh]
   \centering
   \subfloat[Profiles of the azimuthal angle]{
   \resizebox{80mm}{!}{\includegraphics*{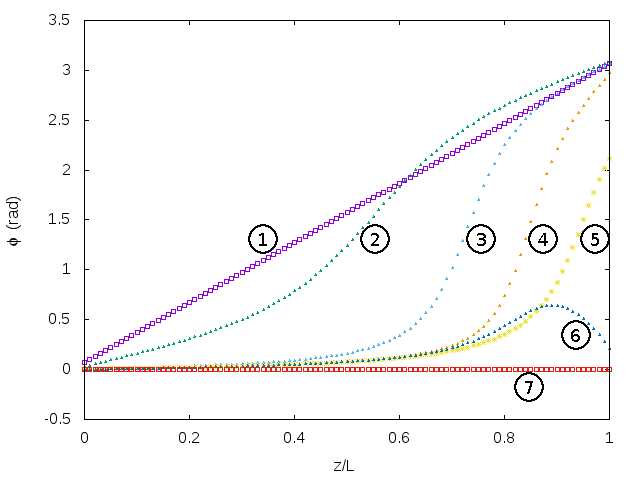}}
   \label{fig:lh_nem_up_phi}
}
   \subfloat[Profiles of the polar angle]{
   \resizebox{80mm}{!}{\includegraphics*{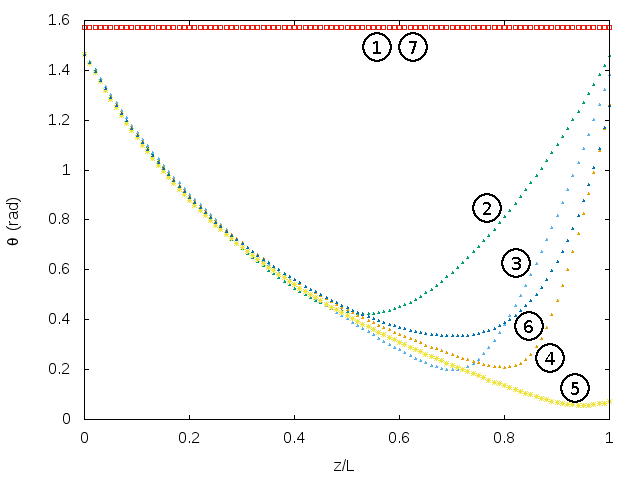}}
   \label{fig:lh_nem_up_theta}
}
\caption{
Profiles of (a)~the azimuthal and (b)~the polar angles for each image in the
  MEP for the anchoring breaking transition.
  }
\label{fig:up_phi_theta}
\end{figure*}

%%%%%%%%%%%%%%%%%
\subsection{Anchoring breaking transitions}
\label{subsec:anch-break-trans}
%%%%%%%%%%%%%%%%%%

In order to study an alternative scenario of the unwinding transition,
we have used
another starting approximation for the MEPs
that involves
the CLC structures where the uniform twist
from $\phi_{-}=0$ to $\phi_{+}$
is superimposed by the out-of-plane director deformation
with the polar angle $\theta$ 
varying from $\theta_{-}=\pi/2$ to $\theta_{+}$.
In this approximation,
similar to the case of the director slippage transitions, 
the twist angle
$\phi_{+}$ along the path monotonically unwinds
changing from $-\pi$ to zero.
By contrast,
for the polar angle $\theta_{+}$ at the top substrate,
the initial decrease from $\theta_{+}=\pi/2$
to the value close to zero $\theta_{+}\approx 0$
that occurs in the first half of the path,
is followed by the increase in $\theta_{+}$
that restores its initial value $\theta_{+}=\pi/2$
at the final state in the second half of the path.
So, the initial guess for the MEP
assumes that, for the transition state,
the CLC director at the top substrate
is nearly normal to the bounding surface.
Such a starting guess will be referred to as
the \textit{anchoring breaking approximation}.

Figure~\ref{fig:energy_mep_up}
presents the results for
the energies of the CLC structures along the  MEP computed at
$U=0.53$~V and $W_{\theta}/W_{\phi}=1.25$
using the initial anchoring breaking approximation.
Similar to Fig.~\ref{fig:energy_mep_rotate},
the first image
is the metastable left-handed helical structure and
the last image
represents the stable untwisted state.
The energy barrier is determined by
the energy of the transition state
corresponding to the fifth numbered image.

As is illustrated in
Fig.~\ref{fig:energy_mep_up},
in the first half of the path,
the region of pronounced out-of-plane deviation approaches
the top boundary surface reaching the transition (saddle-point) structure
with the director orientation close to the normal to the substrate
($\theta \approx 0$).
Such structure implies
that the twisted structure
unwinds via
anchoring breaking
that occurs at the upper substrate
and such unwinding transition will be referred to
as the anchoring breaking transition.

Note that, in contrast to the director slippage transition
(see Fig.~\ref{fig:energy_mep_rotate}),
there is a flatten region in the close vicinity of the maximum
of the energy curve shown in Fig.~\ref{fig:energy_mep_up}.
The reason is that, at $\theta\approx 0$,
variations in the azimuthal angle $\phi$
have negligible effect on the energy.
From the other hand,
these variations produce noticeable changes
in the reaction coordinate which is the normalized sum
of the angles, $\theta$ and $\phi$,
along the path. 

The profiles of the azimuthal and
the polar angles computed for the images of the anchoring breaking
transition shown in Fig.~\ref{fig:energy_mep_up}
are presented in
 Figs.~\ref{fig:lh_nem_up_phi} and~\ref{fig:lh_nem_up_theta},
respectively. 
It can be seen that, except for the initial and final states,
all the profiles are nonlinear.
The fifth curve for the polar angle of
the transition state
clearly shows homeotropic orientation of the director
at the upper substrate with
$\theta_{+}\approx 0$.

\begin{figure*}[!tbh]
  \centering
  \resizebox{100mm}{!}{\includegraphics*{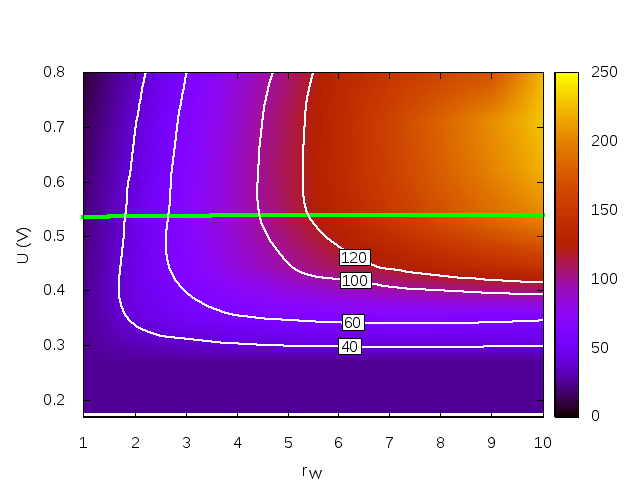}}
  \label{fig:lh_nem_up}
  \caption{Energy barrier map in the $r_W-U$ plane for the anchoring breaking transitions.
    The green line indicates the threshold voltage of the
    Fr\'eedericksz transition $U_{th}$
    the energy barrier scale is
    given in $\mu$J/m$^2$.}
  \label{fig:barrier_map_up}
\end{figure*}

In Fig.~\ref{fig:barrier_map_up},
we show the results for the energy barrier
computed as a function of the anchoring ratio
$r_W$ and the voltage $U$.
From the energy barrier map
presented in Fig.~\ref{fig:barrier_map_up},
it can be inferred that
the energy barrier is independent of $r_W$ and $U$
until the voltage exceeds its critical value
$U_{c}\approx 2.9$~V.
So, at $U<U_c$,
the anchoring breaking approximation produces
the results identical to the director slippage transitions.
For the voltages above $U_c$,
the tilted CLC structures come into play
and the anchoring strength
$W_{\theta}$ will affect the values of
the energy barrier.

\begin{figure*}[!tbh]
   \centering
   \subfloat[$\theta_{\ind{min}}$ for the
   director slippage transition.]{
   \resizebox{80mm}{!}{\includegraphics*{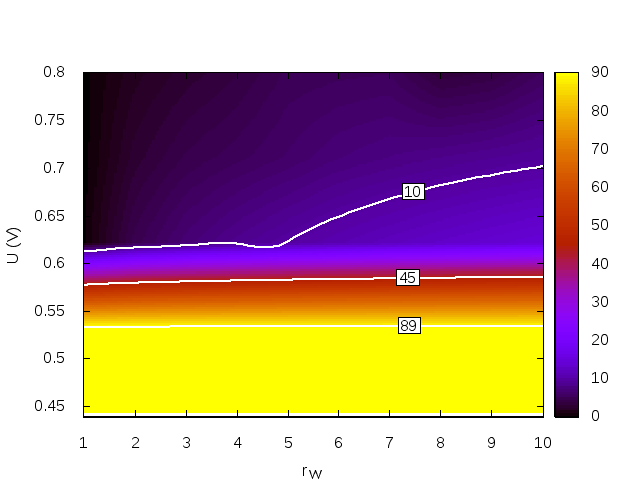}}
   \label{fig:theta_min_saddle_rot}
}
   \subfloat[$\theta_{\ind{min}}$ for the
   anchoring breaking transition.]{
   \resizebox{80mm}{!}{\includegraphics*{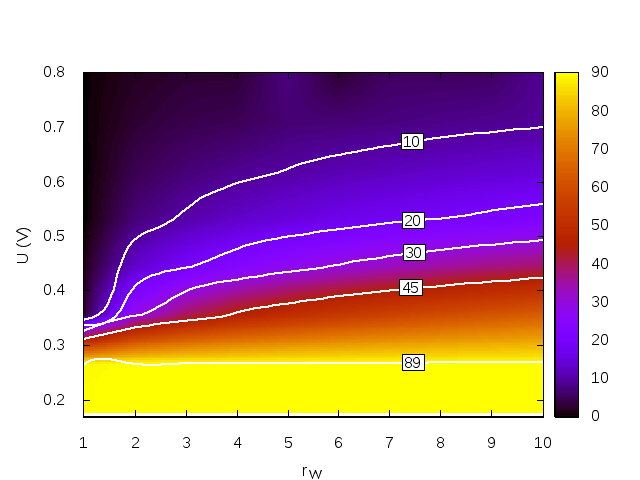}}
   \label{fig:theta_min_saddle_up}
}
\caption{Map of the minimum polar angle
  $\theta_{\ind{min}}$ ($\equiv\min_z\theta(z)$)
  at the saddle point (transition state)
  in the $r_W-U$ plane.}
\label{fig:theta_min_saddle}
\end{figure*}

\subsection{Comparison between two scenarios}
\label{subsec:crossover-trans}

Below the threshold voltage
$U<U_{\ind{th}}$, dependence of the energy barrier
on $W_{\theta}$  originates from the tilted saddle-point structure
representing the transition state.
The corresponding out-of-plane deviations
of the director can be quantitatively described
by the minimum polar angle
\begin{align}
  \label{eq:theta_min}
  \theta_{\ind{min}}=\min_z\theta(z)
\end{align}
evaluated for the transition state of a MEP.
The smaller the value of $\theta_{\ind{min}}$ is
the more tilted the saddle point state is.

The maps of the angle $\theta_{\ind{min}}$
in the $r_W-U$ plane calculated using
the director slippage and the anchoring breaking
initial approximations
are shown in
Figs.~\ref{fig:theta_min_saddle_rot}
and~\ref{fig:theta_min_saddle_up},
respectively.
Referring to Fig.~\ref{fig:theta_min_saddle},
it can be seen that, for the anchoring breaking transitions,
a noticeable decrease in $\theta_{\ind{min}}$
takes place when the voltage exceeds $U_c$,
whereas, for the director slippage transitions,
this happens in the close vicinity of the threshold voltage
$U_{\ind{th}}$.
So, we arrive at the conclusion that
the two mechanisms become essentially distinct
at voltages higher than $U_c$. 

\begin{figure*}[!tbh]
  \centering
  \resizebox{120mm}{!}{\includegraphics*{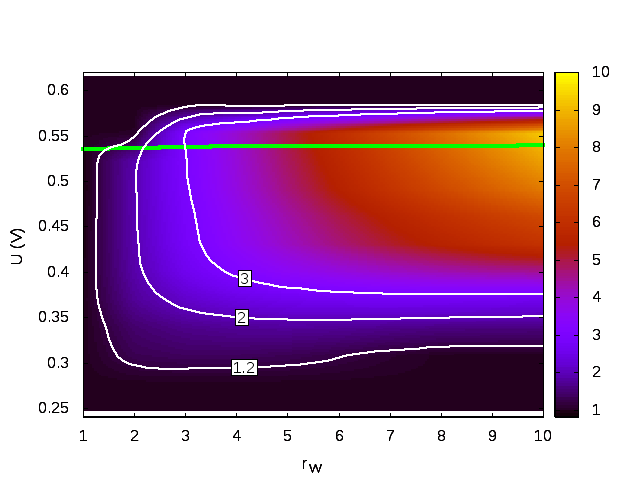}}
\caption{Map of the ratio of the energy barriers for the anchoring
  breaking and the director slippage transitions  in the $r_W-U$
  plane. 
} 
\label{fig:ratio_map}
\end{figure*}

In order to further emphasize the difference between the two
scenarios,
figure~\ref{fig:ratio_map} presents the distribution of the ratio of the energy 
barriers in the $r_W-U$ plane.
It can be seen that, at $U>U_c$,
this ratio is an increasing function of $r_W$
and, when the difference
between the azimuthal and polar anchoring strengths is small
with $r_W\approx 1$, the barriers are nearly equal.

\begin{figure*}[!tbh]
   \centering
   \resizebox{80mm}{!}{\includegraphics*{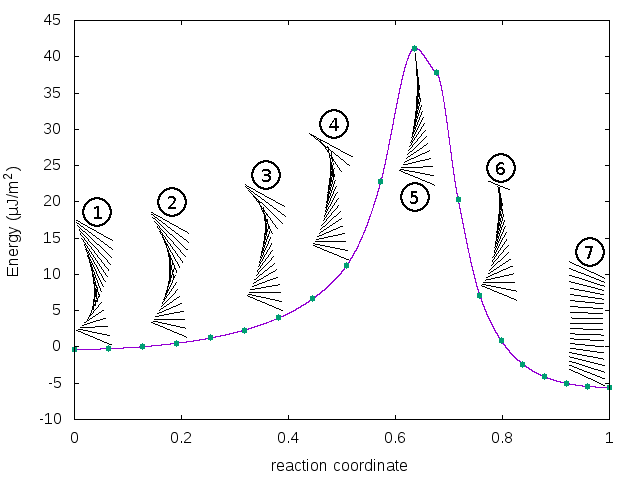}}
   \caption{
  Energy per unit area along the MEP for the transition above the Fr\'eedericksz threshold
     computed at $r_W=2$ and $U=0.64$~V.
   }
\label{fig:energy_mep_064}
\end{figure*}

\begin{figure*}[!tbh]
   \centering
   \subfloat[Profiles of the azimuthal angle]{
   \resizebox{80mm}{!}{\includegraphics*{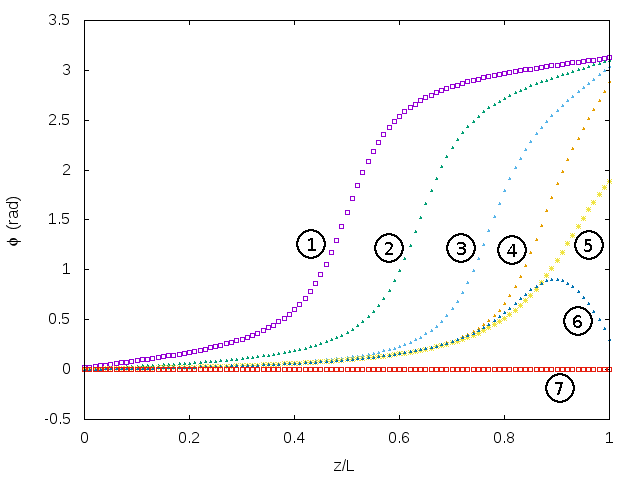}}
   \label{fig:phi_064}
}
   \subfloat[Profiles of the polar angle]{
   \resizebox{80mm}{!}{\includegraphics*{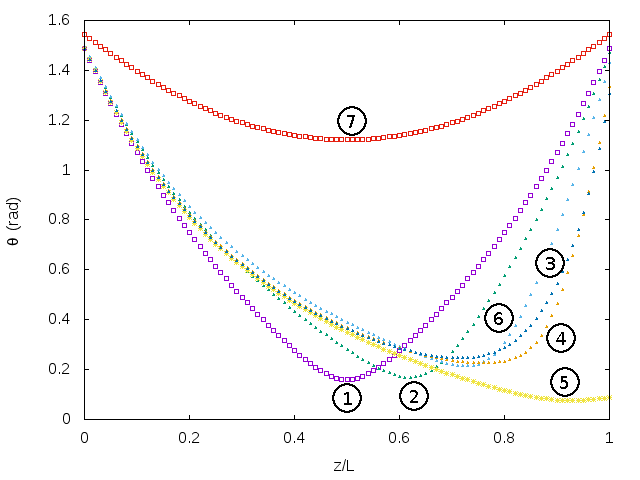}}
   \label{fig:theta_064}
}
\caption{
  Profiles of (a)~the azimuthal and (b)~the polar angles for the images
  along the MEP calculated above the Fr\'eedericksz transition at
  $r_W=2$ and $U=0.64$~V.} 
\label{fig:phi_theta_064}
\end{figure*}

It is turned out that,
even after the Fr\'eedericksz transition takes place,
the barriers of the MEPs may significantly differ.
The maximum value of the barrier ratio
is reached at large values of anchoring strength ratio in the voltage
interval lying just above the threshold voltage.
This is the interval
where, according
to Figs.~\ref{fig:barrier_map_rotate} and~\ref{fig:theta_min_saddle},
the effect of the saddle point tilt
for the director slippage scenario
is much less pronounced as compared to
the MEPs evaluated using the anchoring breaking initial approximation.

When the voltage further increases,
the transition state director structures
will be dominated by
the field induced deformations.
As is clearly demonstrated in
Fig.~\ref{fig:ratio_map},
the result is that the two scenarios eventually merge into the one
and become indistinguishable
at sufficiently voltages above the Fr\'eedericksz threshold.

We have
used the initial anchoring breaking approximation
to compute the MEP in this high-voltage region
($U=0.64$~V) at the anchoring strength ratio $r_W=2$.
The results for the energies along the MEP
and for the profiles of the azimuthal and polar angles
are shown in Figs.~\ref{fig:energy_mep_064}
and~\ref{fig:phi_theta_064}, respectively.
Interestingly, the transition state
(the fifth numbered image and its profiles)
of this MEP bears
close resemblance to the one for the anchoring breaking
transition (see Figs.~\ref{fig:energy_mep_up}
and~\ref{fig:up_phi_theta}).
By contrast to the latter, the initial twisted and the final untwisted
states (the first and the seventh images, respectively)
of this transition reveal significant out-of-plane
deformations induced by the electric field.

%%%%%%%%%%%%%
\section{Conclusion}
\label{sec:conclusion}
%%%%%%%%%%%%%

In this paper, we have studied
the minimum energy paths for the unwinding transition
in the chiral nematic liquid crystal cell.
Such pathways connect the metastable
CLC states which are local minima of the multidimensional free energy surface,
and the energy of the saddle-point (transition state) of the paths gives
the energy barrier separating the metastable states.
Therefore, the MEP and its saddle points characterize
the mechanism (scenario) of the transition.

We have employed
the geodesic nudged elastic band
(GNEB) method as a computational procedure
to evaluate the MEPs. This method
requires an initial guess for the path
and various starting approximations can
generally produce different MEPs.
In our approach,
this dependence of the MEPs on the starting approximation
is exploited to examine two scenarios of the unwinding transition
from the metastable left-handed CLC helix
to the ground untwisted state.
For this purpose,
we have used
the director slippage (see Sec.~\ref{subsec:direct-slipp-trans})
and the anchoring breaking (see Sec.~\ref{subsec:anch-break-trans}) approximations
as the starting paths for the MEPs.

These MEPs and the energy barriers are
calculated at various values of the voltage $U$ applied across the CLC
cell
and the anchoring strength ratio $r_W=W_{\theta}/W_{\phi}$.
For the director slippage scenario,
orientational configuration of the transition state
is found to remain planar until the voltage reaches
the Fr\'eedericksz threshold $U_{\ind{th}}$.
By contrast,
in the saddle-point state of the MEPs representing the anchoring
breaking scenario of the transition,
out-of-plane deformations of the director structure are
localized near the bounding surface chosen by the initial
approximation.
This scenario come into play only when the voltage
exceeds its critical value $U_c<U_{\ind{th}}$.

At voltages above the Fr\'eedericksz threshold $U_{\ind{th}}$,
both the initial metastable twisted state and
the final untwisted state are deformed
by the applied electric field.
In this case, electrically induced deformations
also affect the saddle-point state of
the MEPs computed using the director slippage approximation.
The result is that
the difference between the two scenarios
becomes negligible at sufficiently high voltages.
Note that, in the low-voltage region where $U<U_c$,
the director structure of the transition state is a
planar Grandjean texture~\eqref{eq:clc_helix}
and the MEPs, similar to the high-voltage regime,
are indistinguishable.

We conclude with the remark
that
the saddle-point states can also be found
without knowledge of the final state
using
the minimum mode following method
recently suggested in Ref.~\cite{Muller:prl:2018}.
This promising method
being more complicated than the GNEB method
allows discovering unexpected final states
and the transition mechanisms.

 \begin{acknowledgments}
   This work was supported by Russian Foundation of Basic Research
   under the grant  18-02-00267 and by the Foundation for Advancement
   of Theoretical Physics and Mathematics “BASIS” under
   the grant 19-1-1-12-1.
 \end{acknowledgments}

%\bibliographystyle{apsrev}

%\bibliography{optics,polymer,scatter,lc,quant,hk,flc,qft,math,my_papers}
%\bibliography{tenishch}


\end{document}